\documentclass{article}

\usepackage{PRIMEarxiv}

\usepackage[utf8]{inputenc} % allow utf-8 input
\usepackage[T1]{fontenc}    % use 8-bit T1 fonts
\usepackage{hyperref}       % hyperlinks
\usepackage{url}            % simple URL typesetting
\usepackage{booktabs}       % professional-quality tables
\usepackage{amsfonts}       % blackboard math symbols
\usepackage{nicefrac}       % compact symbols for 1/2, etc.
\usepackage{microtype}      % microtypography
\usepackage{lipsum}
\usepackage{fancyhdr}       % header
\usepackage{graphicx}       % graphics
\graphicspath{{media/}}     % organize your images and other figures under media/ folder

\usepackage{float}
\usepackage{xcolor}

\definecolor{biocolor}{HTML}{45A2BC}
\definecolor{lscolor}{HTML}{FCB515}
\title{Conservation priority mapping to prevent zoonotic spillovers}

\author{
        Leonardo Teixeira Viotti \\
	Department of Biological Sciences\\
	University of Pittsburgh\\
	Pittsburgh, PA - USA \\
	\texttt{leonardoviotti@pitt.edu}  \\
   \And
        Luis Diego Herrera \\
	World Bank Group\\
	Washington, DC - USA\\
	\texttt{lherreragarcia@worldbank.org}
    \And
        Garo Batmanian \\
	Servico Florestal Brasileiro\\
	Brasília, DF - Brazil \\
	\texttt{garo.batmanian@mma.gov.br}
    \And
        Franck Berthe \\
	World Bank Group\\
	Washington, DC - USA\\
	\texttt{fberthe1@worldbank.org}
    \And
        Rachael Kramp \\
	Department of Biological Sciences\\
	University of Pittsburgh\\
	Pittsburgh, PA - USA \\
	\texttt{rachael.kramp@pitt.edu}  
}

\begin{document}
\maketitle

% --------------------------------------------------------------------------------
\begin{abstract}

Diseases originating from wildlife pose a significant threat to global health, causing substantial human and economic losses each year. The transmission of disease from animals to humans occurs at the interface between humans, livestock, and wildlife reservoirs, influenced by abiotic factors and ecological mechanisms. Although evidence suggests that intact ecosystems can reduce transmission, disease prevention has largely been neglected in conservation efforts and remains underfunded compared to mitigation. A major constraint is the lack of reliable, spatially explicit information to guide efforts effectively. Given the increasing rate of new disease emergence in recent decades, accelerated by climate change and biodiversity loss, identifying priority areas for mitigating the risk of disease transmission is more crucial than ever. We present new high-resolution (1 km) maps of priority areas for targeted ecological countermeasures aimed at reducing the likelihood of zoonotic spillover, along with a methodology adaptable to local contexts. Our study compiles data on well-documented risk factors, protection status, forest restoration potential, and opportunity cost of the land to map areas with high potential for cost-effective interventions. We identify low-cost priority areas across 50 countries, including 277,000 km² where environmental restoration could mitigate the risk of zoonotic spillover and 198,000 km² where preventing deforestation could do the same, 95\% of which are not currently under protection. The resulting layers, covering tropical regions globally, are freely available alongside an interactive no-code platform that allows users to adjust parameters and identify priority areas at multiple scales. Ecological countermeasures can be a cost-effective strategy for reducing the emergence of new pathogens; however, our study highlights the extent to which current conservation efforts fall short of this goal.

\end{abstract}

% % keywords can be removed
% % \keywords{First keyword \and Second keyword \and More}

% --------------------------------------------------------------------------------
\section{Introduction}

Emerging infectious diseases (EIDs) of wildlife origin pose risks to global health, poverty reduction, and economic growth. Even when infectious diseases are not fatal, they often deepen poverty, disrupt livelihoods, and undermine food security \cite{berthe2022putting, mphande2016infectious}. The COVID-19 pandemic and the monkeypox outbreak are among the most recent global health threats originating from wildlife \cite{leendertz2025fire, crits2024genetic, liu2024surveillance}, Other examples from recent decades include HIV/AIDS, Nipah virus disease, avian influenza, Ebola virus disease, SARS, and MERS. In 2020, as a consequence of the COVID-19 economic shutdown, the global economy contracted by 4.3\%, amounting to approximately US\$3.6 trillion \cite{berthe2022putting}.

To reduce the number of emerging infectious diseases in human populations, we must address factors driving zoonotic spillover events. A spillover occurs when a pathogen is transmitted from an animal population to humans. These events are not rare; estimates suggest that 60–75\% of emerging human infectious diseases originate from animals \cite{woolhouse2005host, jones2008global}

Spillovers are driven by increasing contact between wildlife disease reservoirs, humans, and livestock. The primary driver of pathogen transmission from wildlife to humans is recognized to be anthropogenic land-use change \cite{patz2004unhealthy, gottdenker2014anthropogenic, becker2019problem}. Land-use change alters the abundance and distribution of wildlife, creating novel inter-species contacts. These contacts facilitate pathogen spread, which can eventually lead to human infection and further transmission \cite{faust2018pathogen}. For example, urbanization and deforestation have changed fruit bat distribution, pushing bats closer to pig farms. Pigs acted as amplifier hosts, leading to Nipah virus spillover to humans in Malaysia in 1998 \cite{kulkarni2013nipah, pulliam2012agricultural}.

Humans interacting with livestock in high densities, particularly pigs and poultry, represent a major risk factor \cite{klous2016human}. Livestock can become intermediate hosts that amplify viral loads and increase the likelihood of successful transmission to humans \cite{pulliam2012agricultural}. Intermediate hosts can also act as “mixing vessels,” in which infection by different viral strains enables recombination or reassortment, generating new strains \cite{ellwanger2021zoonotic}. Strategies to reduce livestock acting as intermediate hosts include lowering animal density, improving sanitary conditions, and ensuring proper PPE and working conditions for animal caretakers \cite{WOAH2022}. While pathogens may exist in reservoir populations without infecting humans, certain ecological processes can increase the likelihood of spillover \cite{plowright2017pathways}.

Intact ecosystems create barriers that prevent pathogens from reaching humans or livestock \cite{plowright2024ecological}. Intact ecosystems can regulate host populations and increase the prevalence of non- or less-competent hosts, making transmission more difficult \cite{civitello2015biodiversity}. Conversely, degraded environments tend to favor resilient species that often host pathogens. These species are more likely to migrate to rural and urban areas, increasing proximity to humans \cite{gibb2020zoonotic}. Habitat changes influence species composition, host abundance, vector dynamics, pathogen life cycles, exposure pathways, and selection pressures. Disturbances can also increase pathogen virulence \cite{myers2009emerging, mcfarlane2013land}. The likelihood of pathogen-human encounters depends on host and human population densities and environmental characteristics. As humans encroach on wildlife habitats, interactions with animals become more frequent, raising spillover risk.

Conservation efforts, such as habitat protection and restoration, can mitigate spillover by reducing human-wildlife and livestock-wildlife interactions, enhancing dilution effects, and lowering reservoir host density \cite{patz2004unhealthy}. Habitat fragmentation and deforestation disrupt predator-prey dynamics, often increasing reservoir abundance, as seen in Lyme disease \cite{patz2004unhealthy, kilpatrick2017lyme}. Preserving intact habitats, such as forests, can reduce transmission by supporting diverse species that limit reservoir dominance or by maintaining remote distributions of high-risk species, as observed in Hendra and Nipah virus systems.

Here, we propose an adaptable methodology to identify low-cost priority areas for targeted conservation at high spatial resolution, with the goal of preventing zoonotic spillovers. We identify areas for both deforestation prevention and restoration of degraded lands. Using openly accessible data on risk factors, including deforestation risk, biodiversity, and livestock density, we present a spatially explicit approach for identifying high-risk areas. This open-source approach can guide conservation targets at multiple scales and highlight areas not covered by existing efforts, which rarely consider disease prevention.

With this methodology, we create high-resolution (1 km) risk maps for tropical regions. Tropical forests account for a large proportion of spillover events, particularly in biodiverse areas undergoing land-use change \cite{allenglobal}. We map two components of spillover risk: one directly from wildlife to humans and another mediated through livestock. Areas where both risks overlap can act as catalysts for new disease emergence, as livestock serve as intermediary hosts; we classify these as priority areas. We divide priority areas by intervention type: deforestation prevention or ecological restoration through natural regeneration. Across 50 countries, we identify 277,000 km² where restoration could mitigate spillover risk and 198,000 km² where deforestation prevention could do the same, 95\% of which are not under protection. This analysis is available in as no-code \href{https://ee-oneh.projects.earthengine.app/view/zspill-map-beta}{Google Earth Engine application}.

Investing in prevention costs a fraction of mitigation \cite{dobson2020ecology}, yet current investments focus on mitigation \cite{berthe2022putting}. Environmental degradation is clearly linked to spillover risk \cite{gibb2020zoonotic, gibb2024anthropogenic, ellwanger2021zoonotic}, and reducing degradation in high-risk areas offers high returns, even without considering co-benefits such as carbon capture or ecosystem services \cite{vora2022want, bernstein2022costs}. However, policymakers lack clear guidance on where and how to invest.

Previous studies have mapped the risk of zoonotic spillovers or emergent infectious diseases of wildlife origin, but with limited guidance on where ecological interventions are feasible or cost-effective. Jones et al. (2008) \cite{jones2008global} analyzed 335 EID events from 1940 to 2004 to identify spatial and temporal patterns and found that events were correlated with socio-economic, environmental, and ecological factors, such as human population density and biodiversity. Allen et al. (2017) \cite{allenglobal} combined historical EID events with spatial predictors to develop a global zoonotic EID index at 1\textdegree resolution ($\approx$100 km). Walsh et al. (2020) \cite{walsh2020whence} mapped human-livestock-wildlife interfaces and weak health systems at $\approx$10 km resolution. Similarly, there has been extensive discussion about areas of conservation priority either for protection or restoration. Notable efforts have focused on factors such as biodiversity hotspots or representation of different ecosystems (e.g., \cite{myers2000biodiversity}, \cite{olson2002global}, and \cite{brooks2006global}) other studies have focused on ecosystem restoration and carbon capture (e.g., \cite{strassburg2020global} and \cite{busch2024cost}) and other ecosystem services such as water and air quality, and pollination (e.g., \cite{naidoo2006mapping}, and \cite{jung2021areas}). However, prevention of disease transmission to humans has been neglected in the literature.

As humans expand their footprint on the planet, encroaching into and altering natural habitats, the potential for zoonotic disease emergence has increased rapidly. The pace of infectious disease emergence has accelerated over the past 70 years. The number of outbreaks has grown from fewer than 100 per year before the 1980s to several hundred since 2000 \cite{morand2020accelerated}. Additionally, disease richness (the number of unique diseases) and disease diversity (richness and outbreak evenness) have increased significantly over this period \cite{smith2014global}. The annual probability of extreme epidemics occurring could increase up to threefold in the coming decades \cite{marani2021intensity}. 

Reactive strategies are insufficient to combat the emergence of new diseases, major outbreaks, or pandemics; they are also extremely costly \cite{dobson2020ecology}. Yet prevention remains largely neglected \cite{berthe2022putting, plowright2024ecological}. While factors such as short-term economic benefits from deforestation play a role, a key limiting factor is the lack of reliable spatially explicit high-resolution information on spillover risk to inform local interventions.

% --------------------------------------------------------------------------------
\section{Data and Methods}

Land conversion and environmental degradation in tropical areas—particularly tropical forests—are considered major drivers of diseases of zoonotic origin \cite{morand2020accelerated}. To highlight areas for conservation priority, we selected spatially available information on well-documented risk factors that cover tropical regions of the globe. We identify two primary layers of transmission risk stemming from livestock and wildlife. The first layer highlights areas with high densities of livestock and humans. The second layer highlights areas with high biodiversity of selected taxa and distinguishes between those under the threat of deforestation and those that are degraded but suitable for restoration. Where these two layers overlap, we identify priority areas—locations with an elevated likelihood of transmission based on the accumulation of risk factors.

\subsection{Data processing}

Data processing and analysis were performed using Google Earth Engine (GEE) \cite{gorelick2017google}, a platform for the analysis of geospatial datasets that is freely available for scientific, academic, and non-profit use. To ensure spatial consistency, all data sources were brought into a common coordinate reference system and aggregated or resampled to a 1 km resolution. Data sources originally at finer resolutions were aggregated into larger pixels, while sources coarser than 1 km were resampled into smaller pixels; that is, a larger pixel is divided into smaller ones that inherit the original value. Pixels that are not perfectly aligned are weighted according to the proportion of their area within the larger pixels. All of the layers are publicly available and can be downloaded from their respective sources. They can be compiled to crate priority areas with the open-source codebase and available in our \href{https://github.com/LeonardoViotti/zspill}{GitHub repository}, where it can be adapted to use other data sources or to fine-tune parameters to address regional priorities. Additionally, we provide a no-code tool that allows users to work with the provided layers and adjust analysis parameters.

\subsection{Data Layers}

To map environmental degradation in tropical areas across the globe, we highlighted areas at risk of deforestation by 2050 and areas suitable for restoration through natural regeneration. For both of these, we also consider the opportunity cost of the land \cite{busch2024cost} in perpetuity, applying a 7\% interest rate on 2020 USD/ha. The restoration suitability layer consists of degraded areas suitable for natural regeneration \cite{griscom2017natural, busch2024cost}. The data are provided at 1 km resolution but were reprojected to match our other layers. We combine the provided layers to identify currently degraded areas suitable for ecological countermeasures.

Deforestation risk comes from Vieilledent et al. (2022) \cite{vieilledent2022spatial}, who use a logistic regression model with spatial random effects to estimate the probability of deforestation across 119 tropical countries, using explanatory variables such as topography, accessibility, forest fragmentation, deforestation history, and protection status. We use one of the datasets provided, containing 30 m pixels assigned values of 0 or 1 for predicted deforestation by 2050. We include in our deforestation risk layer 1-km pixels that had at least 10\% of their area with standing forest in 2020 and a predicted loss of at least 30\% of that standing forest by 2050.

We additionally filtered both deforestation risk and restoration potential areas to sites with an opportunity cost under USD 7,200/ha in perpetuity. The threshold selected is at the high end of commercial timberland values in Low- and Middle-Income Countries \cite{cubbage2020global}. The idea is not to assume this land will be used for commercial timber, but to restrict the prioritization to areas where conservation is likely to be viable and exclude sites where the value for food production and other forms of agriculture is high.

For areas with high biodiversity, we selected species richness of four taxa associated with spillover risk to humans: birds, bats, rodents, and primates \cite{ellwanger2021zoonotic, luis2013comparison, plowright2017pathways}. Data for Aves come from BirdLife International (2020)\footnote{BirdLife International, Data Zone. http://datazone.birdlife.org/species/requestdis downloded on Apr 09 2025}. Data for the mammalian orders Chiroptera, Rodentia, and Primates come from IUCN \cite{jenkins2013global, pimm2014biodiversity}. The original layers are available at a resolution not consistent with our other data. To match the other layers in our analysis, the data were resampled to the 1 km resolution of the deforestation risk data. Areas were classified as “high biodiversity” if they ranked above the 85th percentile globally for species richness in any of the selected taxa.

Livestock density data for pigs and chickens in 2020 are derived from the FAO and described by Gilbert (2018) \cite{gilbert2018global}. This dataset provides the number of animals per km² in global gridded layers at a resolution of 0.0833 decimal degrees—approximately 10 km at the equator. The data were re-projected and resampled to match the higher resolution of the other layers in the analysis, ensuring spatial consistency across layers. Given the long-tailed distribution of livestock density across the globe (see SI), we define “high density” areas as those above the global 97th percentile for animal density (pigs or chickens) within a 10-km pixel.

Human population density for the year 2020 is from the Gridded Population of the World Version 4 (GPWv4) \cite{CIESIN2018_access} and consists of the number of individuals per 30-arc-second grid cell (approximately 1 km). The data were projected to match the other layers, but no resampling or aggregation was applied. Areas were considered high-density if a 1-km pixel contained more than 100 people, which would be classified as medium risk of viral zoonotic spillover by Granger et al. (2020) \cite{grange2021ranking}.

Finally, to identify key protection gaps, we compare the combined biodiversity and livestock risk layer to existing terrestrial or coastal protected areas listed in the World Database on Protected Areas (WDPA) (UNEP-WCMC 2025) \cite{UNEP-WCMC_IUCN_WDPA}, which meet the IUCN definition of a protected area \cite{lopoukhine2012does}.

\section{Results}

\subsection{Conservation priority areas in the tropics}

Tropical regions contain the highest levels of global biodiversity and host a large proportion of the world’s mammalian, avian, and plant species \cite{raven2020distribution, pillay2022tropical}. At the same time, these ecosystems are experiencing some of the fastest rates of land-use change due to agricultural expansion, logging, mining, and infrastructure development \cite{hansen2013high, curtis2018classifying}. This rapid environmental change threatens ecosystem integrity, species persistence, and the stability of ecological processes that can regulate disease dynamics \cite{dirzo2014defaunation}.

Environmental degradation can increase zoonotic spillover risk by restructuring ecological communities and intensifying human–wildlife contact \cite{patz2004unhealthy, plowright2017pathways}. Land-use change selectively favors disturbance-tolerant species, particularly rodents and bats, that exhibit high pathogen competence and fast life histories \cite{han2016global, gibb2020zoonotic}. Spillover events are disproportionately associated with deforestation frontiers and agricultural expansion \cite{allenglobal, faust2018pathogen, plowright2021land}. Habitat fragmentation increases wildlife around human settlements, crops, and waste sites, elevating encounter rates and transmission opportunities \cite{patz2004unhealthy, faust2018pathogen}. In this context, conservation can work as a preventative public-health intervention by maintaining ecological conditions that suppress reservoir dominance and limit pathogen amplification before human exposure occurs \cite{plowright2024ecological}.

While conservation is adopted for the protection of biodiversity and other ecosystem services, it has been systematically neglected for zoonotic spillover prevention \cite{dobson2020ecology, bernstein2022costs}. Limited resources require spatial prioritization to maximize ecological and public-health benefits \cite{wilson2007conserving}. Efforts that protect large, contiguous forest patches are more effective at maintaining trophic complexity and keeping competent host species populations in check than small, isolated fragments \cite{faust2018pathogen, gibb2020zoonotic}. However, most conservation planning has not explicitly incorporated zoonotic risk into site selection \cite{glidden2021human, prist2023promoting}. 

To address this, we identify tropical areas where conservation could reduce spillover likelihood by mapping the spatial overlap of key risk factors. These include (i) areas where high densities of pigs or chickens ($\geq$ 97th global percentile) overlap with human population densities exceeding 100 people per km$ˆ2$, and (ii) regions with high richness of known zoonotic host taxa, primates, rodents, bats, and birds ($\geq$ 85th global percentile), which coincide with environmental degradation.

Environmental degradation is captured using two proxies: projected deforestation risk to 2050 \cite{vieilledent2022spatial} and potential for natural forest regeneration \cite{busch2024cost}. Areas where species richness coincides with deforestation risk are classified as high biodiversity–high deforestation risk, while those overlapping with regeneration potential are classified as high biodiversity–high restoration potential.

\subsection{Priority areas for restoration}

Where feasible, restoration through natural regeneration can function as a public-health intervention. Beyond well-documented co-benefits such as carbon sequestration, biodiversity recovery, and the regulation of hydrological cycles \cite{benayas2009enhancement, chazdon2016carbon}, natural regeneration can re-establish the ecological processes that regulate host populations and pathogen transmission. Degraded landscapes tend to favor disturbance-tolerant species that are disproportionately competent reservoirs for zoonotic pathogens \cite{keesing2021impacts}. Restoration increases structural complexity, promotes resource partitioning, and rebuilds trophic interactions, all of which can suppress the dominance of high-risk hosts by enhancing predation and competition \cite{loch2020recovering, bullock2022future}. Finally, restoration can reduce edge density and shrink the human–wildlife interface, lowering encounter rates between people, livestock, and disease reservoirs and thereby reducing spillover risk \cite{mancini2024landscape}.

To provide priority areas for conservation in the tropics, we identify low-cost regions across 50 countries, with 277,000km2 where environmental restoration could mitigate the risk of zoonotic spillovers. These are areas suitable for restoration through natural regeneration is \cite{busch2024cost}, where high densities of livestock and humans intersect high biodiversity, and would cost less than U\$7200/ha, by combining the opportunity cost of the land in perpetuity and implementation costs. What most of these areas have in common is the overlap between farming, particularly large-scale livestock operation meets forest fragments with high species richness. Some examples are southeastern Brazil, the Pacific coast of Central America, central and western India, central Ethiopia, and parts of southeast Asia, such as the eastern part of the Malay peninsula, parts of Sumatra, and Java.

\begin{figure}[h]
\label{fig-rest}
\caption{Priority areas restoration through natural regeneration}
\includegraphics[width=16.5cm]{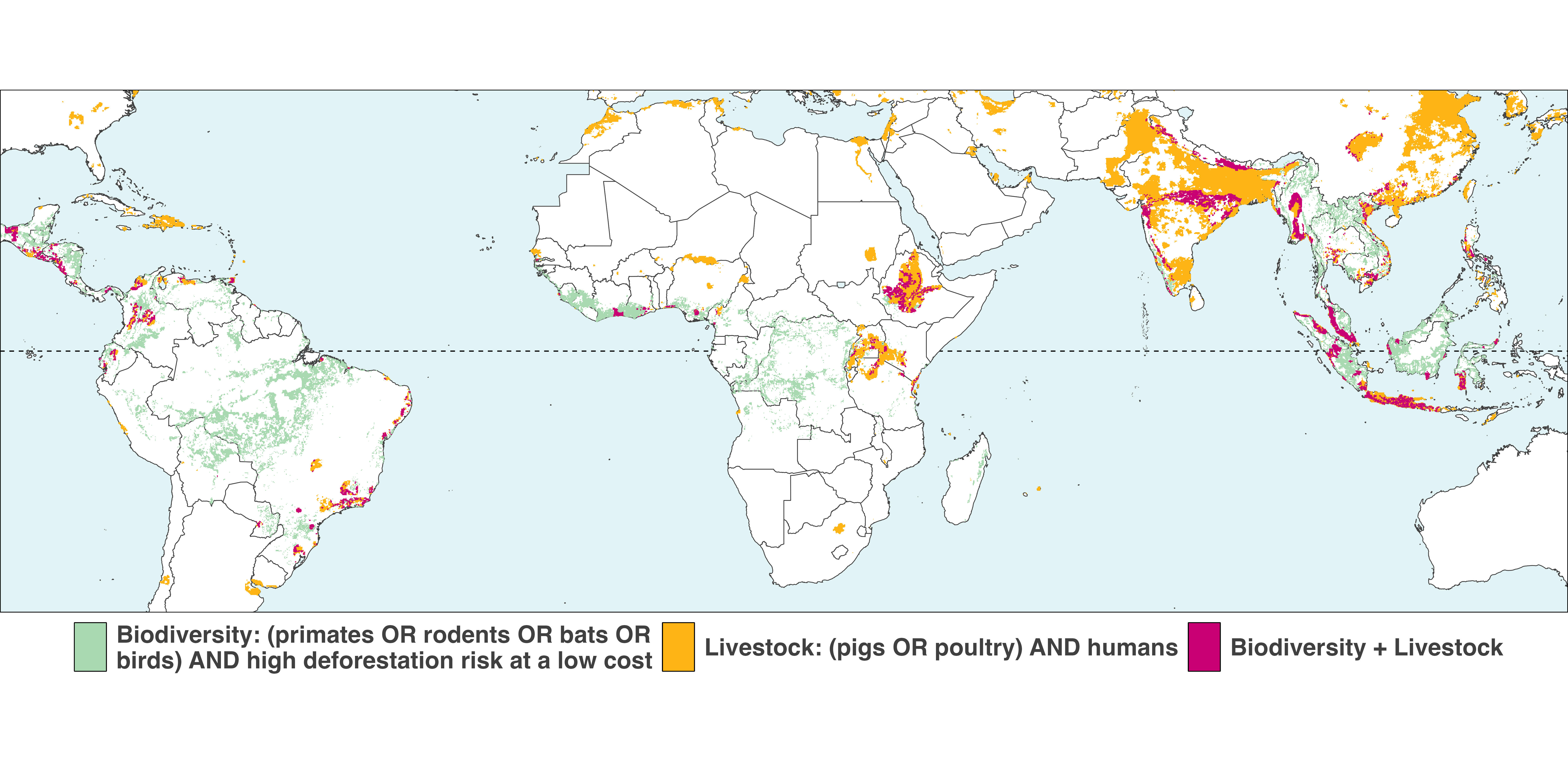}
\end{figure}

\subsection{Priority areas for deforestation prevention}

While ecosystem degradation can be partially reversed through restoration, preventing deforestation in the first place is likely to be the most effective strategy for reducing zoonotic spillover \cite{vora2023interventions}. Land-use change is consistently identified as one of the primary drivers of disease transmission between wildlife and human populations \cite{patz2004unhealthy, faust2018pathogen, plowright2024ecological}. Deforestation fundamentally alters ecological communities and human–wildlife interfaces, creating novel contact zones that facilitate cross-species transmission. As a result, landscapes undergoing rapid forest loss repeatedly emerge as hotspots for infectious disease emergence \cite{patz2004unhealthy, faust2018pathogen, plowright2021land}. 

Disturbance-tolerant species proliferate in agricultural mosaics, forest edges, and human settlements created by deforestation, increasing both host density and contact rates with people \cite{gibb2020zoonotic}. In contrast, intact forests retain trophic complexity and competitive interactions that limit the dominance of these high-risk reservoir hosts \cite{guegan2020forests}. By preventing deforestation, ecosystems can maintain integrity, reducing pathogen amplification and stabilizing transmission dynamics within wildlife communities \cite{vora2023interventions, plowright2024ecological}.

Forest frontiers, which are repeatedly identified as zones of elevated transmission risk \cite{allenglobal, guegan2020forests}. Forest edges created by logging and agricultural expansion concentrate wildlife around crops, waste, and water sources, increasing opportunities for cross-species transmission \cite{patz2004unhealthy}. Continuous forest cover limits these high-risk interfaces by spatially separating human activities from wildlife reservoirs and stabilizing animal movement patterns \cite{plowright2017pathways}. Additionally, deforestation paves the way for land uses that further increase human and livestock densities near degraded habitats, compounding spillover risk through repeated and intensified contact \cite{patz2004unhealthy, plowright2021land}.

To map priority areas for deforestation prevention, we combine spatial layers of high livestock density and human population with regions of high biodiversity currently under threat of forest loss. Where these layers overlap, preventing deforestation could mitigate zoonotic spillover risk. We also filter priority areas to locations where the opportunity cost of land does not exceed 7,200 USD per hectare in perpetuity. We identify approximately 198,000 km² of priority conservation areas across 44 countries where deforestation prevention could yield both biodiversity and public-health benefits.

\begin{figure}[h]
\label{fig-rest}
\caption{Priority areas for deforestation prevention}
\includegraphics[width=16.5cm]{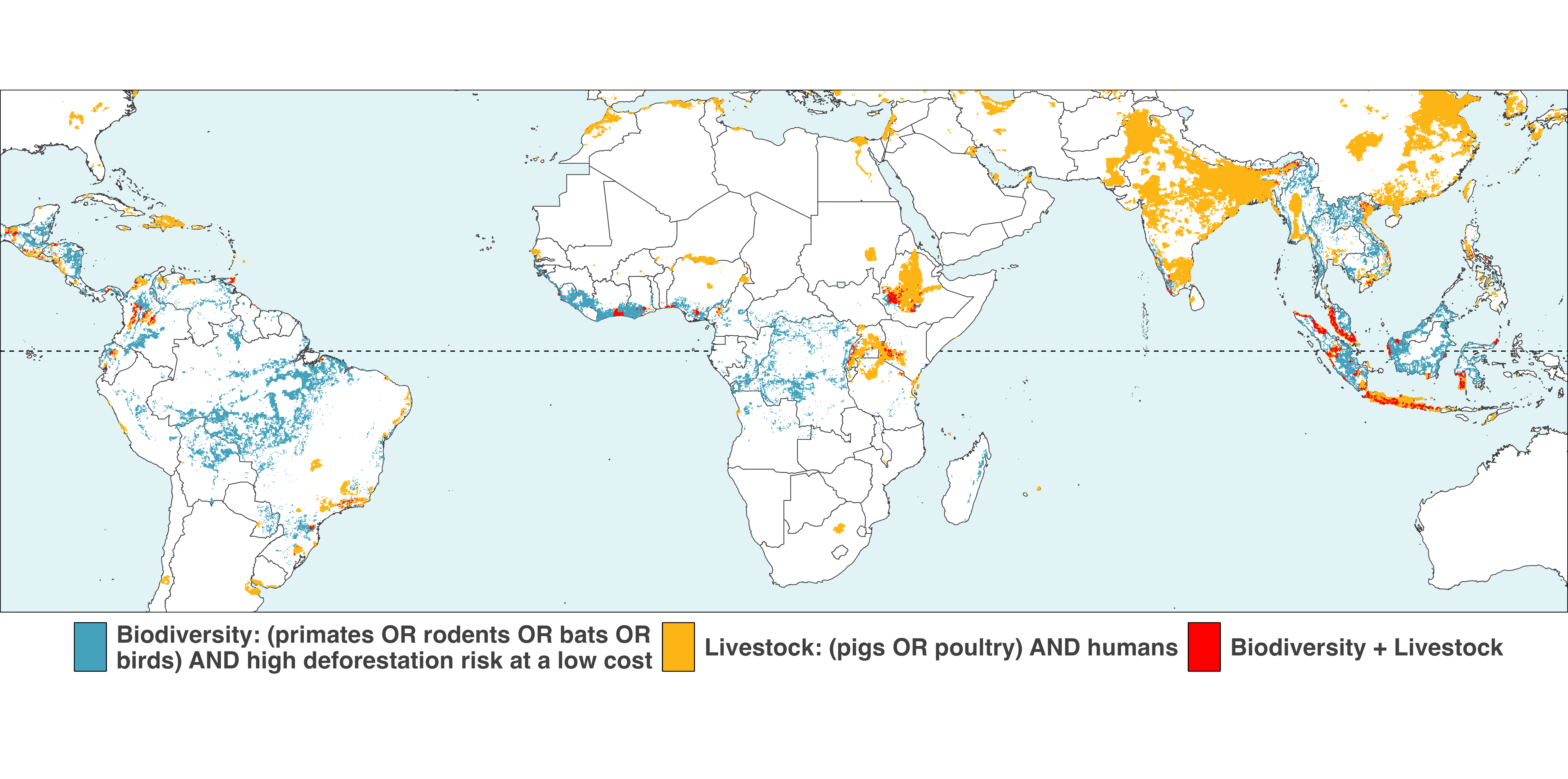}
\end{figure}

Additionally, we identify priority areas that currently fall outside existing protected area networks \cite{UNEP-WCMC_IUCN_WDPA}. Our analysis indicates that approximately 189,000 km², nearly 95\% of all identified priority areas for deforestation prevention, are not covered by formal protection. This finding highlights a substantial mismatch between regions of elevated zoonotic spillover risk and the current global configuration of protected areas. Although protected area coverage has expanded globally over recent decades \cite{jenkins2009expansion}, designations have often been driven by other considerations rather than public-health objectives \cite{terraube2017role}, leaving many high-risk landscapes unprotected.

While the presence of a protected area is an imperfect proxy for effective conservation, it remains a critical first step in limiting land-use change and reducing human encroachment into wildlife habitat \cite{watson2014performance, geldmann2019global}. Even partially enforced protected areas can slow deforestation rates and reduce habitat fragmentation compared to unprotected lands \cite{andam2008measuring}. By restricting agricultural expansion, extractive activities, and infrastructure development, protected areas help maintain ecological integrity and reduce the high-contact interfaces that facilitate spillover events. 

Expanding protected area networks into regions of elevated zoonotic risk could be an integral component of pandemic prevention strategies. Targeting protection toward deforestation frontiers and high-risk human–wildlife interfaces could complement traditional biodiversity goals with explicit public-health benefits

\begin{figure}[h]
\label{fig-rest}
\caption{Priority areas for deforestation prevention outside of protected areas. A Colomnia, B Vietnam, C West Africa...}
\includegraphics[width=16.5cm]{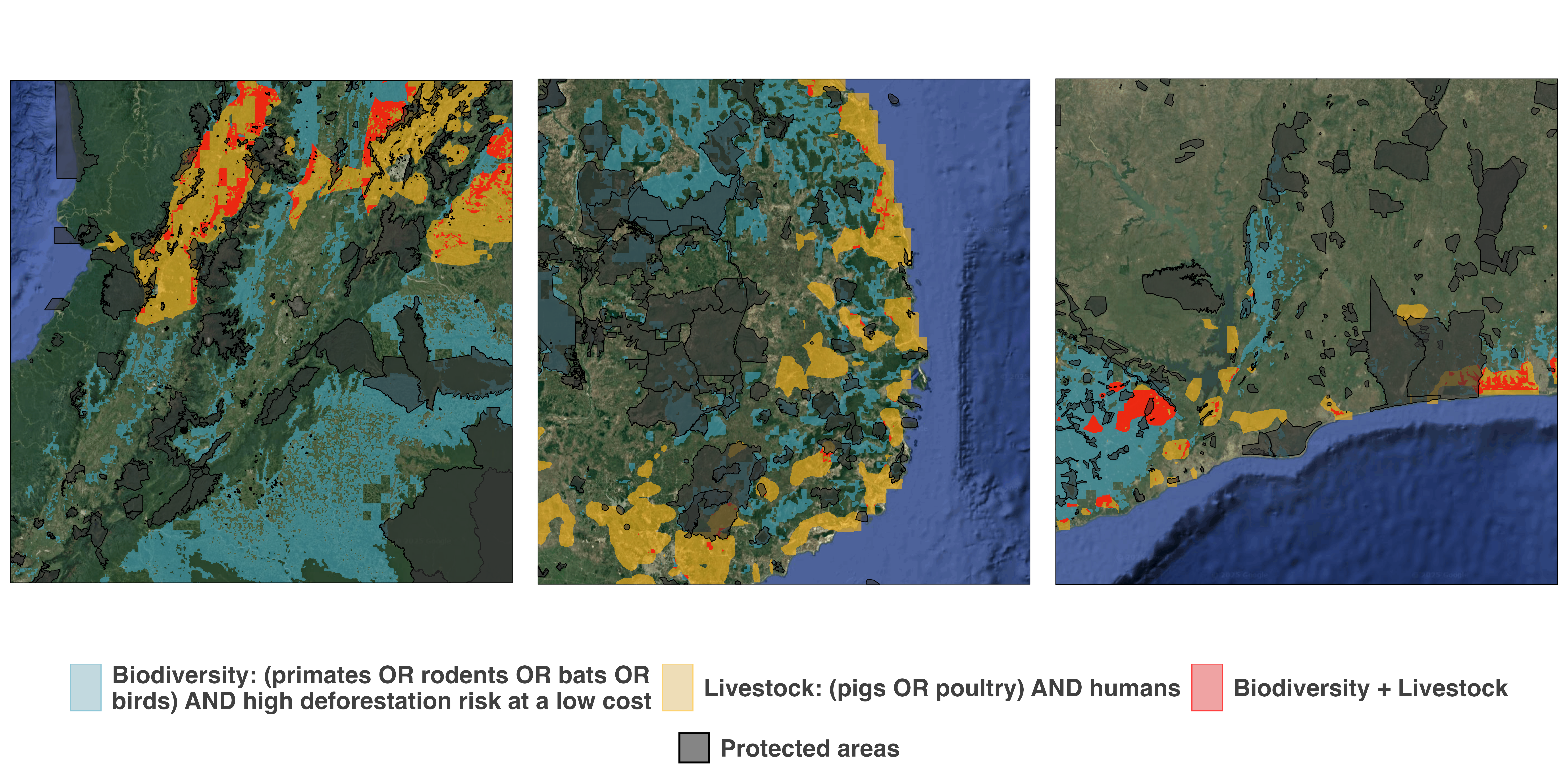}
\end{figure}

Since deforestation prevention and restoration are not mutually exclusive, we further identify a small percentage of total priority areas (approximately 9\%), or 44 thousand km2, where both interventions seem suitable. This is because data sources are independent and use different methodologies, but also because we set a relatively low threshold for standing forest at risk. If at least 10\% of the area of a 1km2 pixel is covered by forest in 2020, it can be included in the analysis. It is perfectly possible for pixels with some area of forest cover flagged for being at risk of deforestation to have other areas suitable for restoration through natural regeneration.

\section{Discussion}

We provide the first high-resolution layers of prioritization for two distinct possible interventions: deforestation prevention and restoration through natural regeneration. We build on previous research that maps spillover risk factors and conservation priorities. Our main contribution is to combine these two approaches into a single framework to prioritize conservation efforts with the specific goal of disease prevention. Our method identifies target areas in tropical regions worldwide for interventions at relatively low cost.

\subsection{Prior work}

Our study aligns closely with previous work to map spillover risk. The methodology presented is based on the mechanistic understanding of how spillovers can occur and what factors can make it more or less likely. Despite not directly incorporating past spillover events in our analysis, our results are very similar to those of studies that did. Areas with both human population density and high biodiversity have been flagged both in our study and other efforts to predict the occurrence of future spillover events based on past occurrences \cite{jones2008global, allenglobal}. Notable areas with considerable overlap include Southeastern Brazil, Sumatra, West African forests, Western India, Andean regions of Colombia, and Forested areas of East Africa.

The identified priority areas also have considerable overlap with other conservation prioritization criteria. A large proportion of our priority areas for both restoration and deforestation prevention are within Biodiversity Hotspots \cite{myers2000biodiversity}. Other large regions have significant areas flagged both in our study and global priority areas for ecosystem restoration \cite{strassburg2020global} and priorities for conserving biodiversity, carbon storage, and ensuring clean water \cite{jung2021areas}. Some notable examples are the Brazilian Atlantic Forest, the Andean regions of Colombia and Peru, West-African forests, the Western Ghats in India, the Malay Peninsula, Sumatra, and Java.

\subsection{Scale and local context}

Although our analysis operates at a pan-continental scale, the implementation of conservation interventions should be done at the local scale. The objective of this study is not to be prescriptive, but to flag regions where further investigation can lead to targeted, context-specific efforts. Ecological outcomes of conservation depend strongly on local conditions, including landscape configuration, species composition, and socioeconomic constraints. Each spatial layer used in our analysis represents a snapshot in time, and the necessary rasterization of complex variables inevitably leads to simplification, potentially introducing bias and uncertainty. Nevertheless, this approach provides a tool that can guide more detailed, site-specific assessments. Adapting our framework to incorporate locally relevant data, such as fine-scale habitat connectivity, hunting pressure, or community land tenure, can guide more effective conservation strategies.

While maintaining healthy ecosystems and reducing contact between humans and wildlife are broadly supported prevention strategies \cite{plowright2024ecological}, the effects of biodiversity on disease dynamics are context-specific \cite{wood2014does}.  Evidence from multiple systems supports the dilution effect, whereby increased biodiversity reduces disease risk by increasing the abundance of incompetent hosts and lowering transmission efficiency \cite{schmidt2001biodiversity, civitello2015biodiversity, keesing2021impacts}. Classic examples include Lyme disease in the northeastern United States, where higher vertebrate diversity reduces the relative abundance of highly competent reservoir hosts. However, the dilution effect is not a universal rule. In certain systems, higher biodiversity may species may lead to more competent hosts or increase vector abundance, resulting in neutral or even amplifying effects on disease transmission \cite{salkeld2013meta, wood2014does}.

How restoration might affect disease transmission, in particular, depends on the ecological context and landscape configuration, as well as community composition \cite{keesing2025ecosystem}. For example, increasing connectivity of fragmented landscapes could lead to high edge density ratios and increase wildlife activity near human settlements and agricultural fields, increasing contact rates and spillover risk  \cite{prist2023promoting}. Effective restoration must therefore prioritize the expansion of interior forest habitat and the reduction of edge effects. 

Response to restoration could be further shaped by species life-history traits. Fast-lived species, with early reproduction, short lifespans, and high population turnover, are typically disproportionately competent hosts that tend to dominate disturbed and early-successional habitats \cite{keesing2021impacts, gibb2020zoonotic}. In contrast, slow-lived species typically exhibit lower competence and are more vulnerable to disturbance. Restoration does not automatically lead to the return of species lost to habitat degradation. Dispersal limitation, local extirpations, hunting pressure, and altered abiotic conditions can prevent the recolonization of slow-recovering taxa even decades after restoration begins \cite{brudvig2017toward}. As a result, restored systems may remain dominated by generalist species unless active interventions, such as assisted migration, reintroductions, or long-term protection, are implemented to rebuild functional communities \cite{volis2019conservation, rummell2024restoration}.

Successful restoration must therefore be tailored to disease-specific transmission pathways, appropriate spatial scales, and local socioeconomic conditions. Communities are affected both by zoonotic risk and by the economic drivers of land-use change, making it essential that restoration strategies balance public health objectives with livelihoods and governance realities. In this sense, restoration is not a one-size-fits-all solution but a powerful ecological tool whose effectiveness depends on careful, context-aware design.

\section{Conclusion}

Our results underscore the extent to which existing conservation efforts fall short of addressing the needs of human health. Although targeted ecological interventions have been shown to reduce the risk of emergence of novel pathogens \cite{plowright2024ecological}, disease prevention remains largely neglected in conservation planning and chronically underfunded relative to outbreak response and mitigation \cite{dobson2020ecology, bernstein2022costs}. This imbalance persists despite growing evidence that prevention is more cost-effective than mitigation \cite{dobson2020ecology, berthe2022putting}. Conservation strategies have historically prioritized factors such as biodiversity protection and carbon storage, while their potential role in pandemic prevention has received comparatively little attention. By explicitly identifying areas where disease transmission risk factors and economic feasibility intersect, our framework helps address this gap and provides a mechanism for integrating public-health objectives into conservation planning.

Ecosystem restoration and deforestation prevention offer viable strategies for pandemic prevention by reducing zoonotic spillover risk. Restoration can rebuild ecological processes that suppress high-risk reservoir species, while deforestation prevention maintains intact ecosystems that buffer human activity from wildlife reservoirs. Together, these strategies address distinct stages of land-use change, from safeguarding remaining intact habitats to repairing already degraded landscapes. However, the effectiveness of ecological countermeasures depends on local landscape configuration, species composition, and socioeconomic context. Our methodology is therefore designed to be adaptable, allow the integration of locally relevant data, and refine priorities according to local realities.

This study highlights priority areas across the tropics where conservation interventions could both reduce the risk of zoonotic spillover and be economically feasible. We flagged of over 277km$ˆ2$ where we believe restoration could reduce the risk of disease transmission from wildlife to livestock and humans, and 197km² where the same could be done by deforestation prevention, 95\% of which lie outside protected areas. This reveals a substantial opportunity to realign conservation investments toward disease prevention. 

By using publicly available spatial layers and providing an interactive platform for exploration, hope to lower barriers for policymakers, practitioners, and local stakeholders to incorporate health considerations into land-use decisions. This approach responds directly to calls for transdisciplinary collaboration between ecology, public health, and development sectors \cite{patz2004unhealthy, tajudeen2022zoonotic}.

Finally, by providing fine-scale, spatially explicit data, our tool supports the integration of ecological research into public-health planning. Biodiversity-focused strategies, particularly those that maintain large, connected habitats and functional trophic structures, can offer a sustainable and cost-effective approach to reducing spillover risk while delivering co-benefits for climate mitigation, food security, and ecosystem resilience. As the frequency of emerging infectious diseases continues to rise under climate change and accelerating land-use conversion, ecological countermeasures should be recognized as essential components of global health security. Reframing conservation not only as a tool for protecting nature but also as a frontline defense against pandemics represents a critical shift toward proactive, preventative public health policy.

% \newpage

%Bibliography
\bibliographystyle{unsrt}

\bibliography{references}

% --------------------------------------------------------------------------------
% \section*{Supplementary materials}

\end{document}